\documentclass[final,5p,times,authoryear]{elsarticle}

\usepackage{amssymb}
\usepackage{amsmath}
\usepackage{float}
\usepackage{multicol}
\usepackage{pgfplots}
\usepackage[separate-uncertainty=true, multi-part-units=single]{siunitx}
\usepackage{dblfloatfix}
\usepackage[dvipsnames]{xcolor}

\usepackage{caption}
\usepackage{subcaption}
\usepackage{graphicx}
\usepackage{booktabs}

\usepackage{url}

\begin{document}

\begin{frontmatter}

\title{Crossover Frequency as a Model-Independent Viscoelastic Constant for Soft Tissue Biomechanics} 

\author[1]{Laura Ruhland\corref{cor1}}
\ead{laura.ruhland@fau.de}
\author[2]{Jing Guo}
\author[2]{Ingolf Sack}
\author[1]{Kai Willner}
\ead{kai.willner@fau.de}
\cortext[cor1]{Corresponding author}
\address[1]{Institute of Applied Mechanics, Friedrich-Alexander-Universität Erlangen-Nürnberg, Egerlandstr. 5, 91058, Erlangen, Germany}
\address[2]{Department of Radiology, Charité - Universitätsmedizin Berlin, Charitéplatz 1, 10117 Berlin, Germany}

\begin{abstract}
Magnetic resonance elastography (MRE) and related elastography techniques are emerging as quantitative diagnostic tools for assessing tissue microstructure and pathology. To determine descriptive parameters of the tissues' properties, a frequency-dependent viscoelastic material model is required, which is calibrated to the measured response in a parameter identification process. However, the selection of this model and the fitting strategy is challenging, since it may influence the identified viscoelastic parameters notably. Here, we address this limitation by proposing the crossover frequency ($f_c$, defined as the frequency at which storage and loss moduli intersect $G'(f_c)=G''(f_c))$ as a model-independent viscoelastic constant for soft tissues. Fresh porcine specimens of the corona radiata, the putamen, the thalamus, and the liver were investigated using tabletop MRE and the frequency-dependent viscoelasticity was characterized with a fractional Kelvin-Voigt model. By validating the crossover frequency against the viscoelastic parameters, we demonstrated that the crossover frequency accurately reflects the viscoelastic behavior, independent of the material model or the fitting strategy. Across all samples, $f_c$ distinguished brain regions and separated brain from liver tissue by median frequencies of \SI{85}{\hertz} (\SI{95}{\percent}  CI: \qtyrange[range-phrase = {-}, range-units = single, number-unit-product = {}]{69}{269}{\hertz})  in the corona radiata, \SI{423}{\hertz} (\SI{95}{\percent}  CI: \qtyrange[range-phrase = {-}, range-units = single, number-unit-product = {}]{316}{575}{\hertz}) in the putamen, \SI{426}{\hertz} (\SI{95}{\percent}  CI: \qtyrange[range-phrase = {-}, range-units = single, number-unit-product = {}]{302}{601}{\hertz}) in the thalamus and \SI{1174}{\hertz} (\SI{95}{\percent}  CI: \qtyrange[range-phrase = {-}, range-units = single, number-unit-product = {}]{1074}{1300}{\hertz}) in the liver (\textit{p}<0.001). These results suggest that crossover frequencies capture distinct viscoelastic fingerprints without requiring viscoelastic model selection. The crossover frequency may therefore serve as a practical, model-independent biomaterial constant to improve comparability of viscoelastic measurements across elastography studies. 

\end{abstract}

\begin{graphicalabstract}
\includegraphics{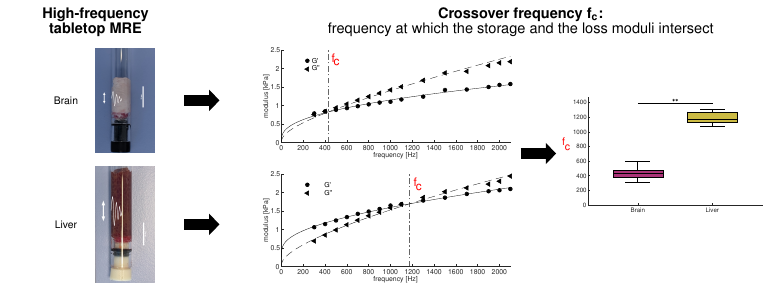}
\end{graphicalabstract}

\begin{keyword}

Magnetic resonance elastography \sep Brain \sep Liver \sep  Viscoelasticity \sep Crossover frequency

\end{keyword}

\end{frontmatter}

\section{Introduction}
\label{Intro}
For the characterization of the tissue microstructure and pathology, magnetic resonance elastography (MRE) and related elastography techniques are advancing as diagnostic tool. MRE is a non-invasive imaging technique, that measures the material properties of soft tissues by imaging the propagation of mechanical waves in the tissue \citep{muthupillai_magnetic_1995}. In the brain, MRE has already been used to study stiffness changes in the in-vivo brain resulting from Parkinson's disease \citep{lipp_cerebral_2013}, Alzheimer’s disease \citep{murphy_decreased_2011} or tumors \citep{janas_vivo_2024}. Another organ which is intensely studied with MRE is the liver. With MRE it is possible to accurately capture different stages of liver fibrosis \citep{asbach_viscoelasticity-based_2010} and MRE is therefore already used in clinical applications for the detection of liver fibrosis \citep{venkatesh_magnetic_2013}. \\

In multifrequency MRE, the mechanical properties of tissue are evaluated at various driving frequencies to measure the frequency-dependent viscoelastic response. Preselected viscoelastic models can subsequently be calibrated to the measured response to obtain frequency-independent material parameters. These parameters are increasingly investigated as potential bio\-markers for tissue properties \citep{manduca_mr_2021}. For the characterization of soft tissues, commonly used models with two material parameters are the Maxwell, the Kelvin-Voigt, and the Springpot model. Of the two parameter models, the Springpot model is the most suitable for biological tissue \citep{sack_impact_2009}. Models with three or more material parameters, such as the Zener, the fractional Zener or the fractional Kelvin-Voigt model, may achieve a stronger agreement with the measured tissue responses then the springpot model. However, a higher number of materiel parameters also results in a greater variability of determined values \citep{klatt_noninvasive_2007}. Since viscoelastic properties reported in the literature vary notably depending on the selected material model \citep{klatt_noninvasive_2007, sack_impact_2009, weickenmeier_brain_2018}, the model choice certainly influences the viscoelastic parameters. A further factor, influencing the resulting parameters, may be the selection of the fitting strategy for the model calibration. \\

To overcome this limitation, we introduce the crossover frequency $f_c$ as a model-independent parameter for viscoelastic tissue characterization. For a viscoelastic solid, the crossover frequency is given by the frequency at which the loss modulus exceeds the storage modulus, and thus the material changes from an elasticity-dominated to a viscosity-dominated behavior. We examined the high frequency response from \SIrange{300}{2100}{\hertz} of three distinct porcine brain regions, i.e. the corona radiata, the putamen and the thalamus, and porcine liver with \textit{ex vivo} tabletop MRE and determined the crossover frequency for every tissue. By validating the crossover frequency against the viscoelastic parameters obtained from calibrating a fractional Kelvin-Voigt model to the measured dynamic moduli, we showed that the crossover frequency reflects the viscoelastic behavior and distinguishes the behavior of the different brain regions as well as the behavior of brain and liver.

\section{Methods}
\subsection{Tissue preparation}
\begin{figure}[h]
    \centering
    \includegraphics{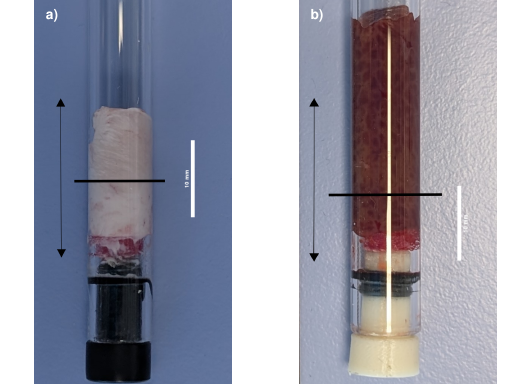}
    \caption{Exemplary tabletop MRE samples of a) brain and b) liver tissues. The black bar demonstrates the position of the image slice, the arrows indicates the vibration direction and the white bars serve as a length scale with a length of \SI{10}{\milli\meter},}
    \label{fig:Sample}
\end{figure}

Our experiments were performed on freshly harvested porcine brain and liver tissue, provided by the local slaughterhouse, which allows us to avoid animal studies for this investigation. At the tabletop MRE the tissue is examined in glass tubes with a outer diameter of \SI{9}{\milli \meter}, an inner diameter of \SI{7}{\milli \meter} and a length of \SI{200}{\milli \meter}. The measurements on brain tissue were taken from our earlier publication \citep{ruhland_combining_2026}. In brief, we tested three different regions from the brain tissue, i.e. the corona radiata (CR, n = 9), the putamen (P, n = 5) and the thalamus (T, n = 5). The cylindrical samples were extracted from the tissue with an \SI{8}{\milli \meter} punch, resulting in a sample diameter of \SI{7}{\milli \meter}, to account for the shape changes of the tissue during the punching process and transferred to the glass tube using a spatula. We only extracted a single sample from every brain, thus each measurement is performed on a new brain and we in total examined 19 brains. For the liver tissue (L, n = 5), the punching process used for the preparation of the brain samples was not applicable, because of the liver's tissue structure. Consequently, we cut samples with dimensions equivalent to those of the brain samples from the liver tissue with a scalpel. For the liver measurements, we used three different porcine livers and extracted one to two samples from each liver. Fig. \ref{fig:Sample} shows exemplary brain and liver sample. We performed the measurement on brain tissue between \SIrange{2}{3.5}{\hour} post-mortem and on liver tissue between \SIrange{3}{5}{\hour} post-mortem. Before the measurement, we stored the tissue in the fridge at \SI{4}{\degreeCelsius} and moistened it with Dulbecco’s
phosphate-buffered saline solution. We extracted the samples from the tissue directly before the measurement.

\subsection{MRE measurements}
A tabletop MRE system, equivalent to the system described in \citet{braun_compact_2018}, was used to measure the tissue's frequency response. A \SI{0.5}{\tesla} permanent magnet MRI scanner (Pure Devices GmbH, Würzburg, Germany) was extended with an external gradient amplifier (DC 600, Pure Devices GmbH, Würzburg, Germany) and a MRI system-controlled piezo-electric driver (PAHL 60/20, Piezosystem Jena GmbH, Jena Germany), to induce shear waves at a certain frequency into the sample. 
We measured the tissue's vibration response at 15 frequencies points between \SIrange{300}{2100}{\hertz} with shear wave amplitudes between \SIrange{2.5}{6.6}{\micro\meter}. The wave images were acquired at four time steps per frequency using a spin-echo MRE sequence. For a single frequency, the acquisition time was \SI{161 \pm 7.2}{\second} for the corona radiata, \SI{181.2 \pm 20.1}{\second} for the putamen, \SI{177.7 \pm 11.9}{\second} for the thalamus and \SI{75.7 \pm 2.4}{\second} for the liver. The repetition times were TR\textsubscript{CR} = \SI{558.7\pm 86.4}{\milli\second}, TR\textsubscript{P} = \SI{639.2\pm 77.1}{\milli\second}, TR\textsubscript{T} = \SI{623.3\pm 46.9}{\milli\second} and TR\textsubscript{L} = \SI{231.8\pm 9.9}{\milli\second}. Further imaging parameters were the same for all tissue types:  echo time TE = \num{27.8} - \SI{29.6}{\milli\second}, slice thickness: \SI{3}{\milli\meter}, field of view: \numproduct{9.6 x 9.6} \si{\milli\square\meter}, matrix size: \numproduct{64 x 64} and resolution: \numproduct{0.15 x 0.15 x 3} \si{\milli\cubic\meter}. All tests were conducted at \SI{37}{\degreeCelsius} to mimic the in vivo conditions. The MRE data acquisition is as well described in detail in \citet{braun_compact_2018}.

\subsection{Viscoelastic modeling}\label{sec:MREDataAnalysis}
To determine the frequency dependent storage and loss modulus, an analytical solution of the shear wave based on the Bessel function was fitting to the measured shear wave at every frequency \citep{braun_compact_2018}. Supplementary Fig. S1 to S3 provides the shear waves and the corresponding Bessel function fits of a representative liver sample. Subsequently, a viscoelastic model was calibrated to the dynamic moduli to obtain the frequency independent mechanical parameters. In this study, we use a two-term fractional Kelvin-Voigt model to describe the material response. This model, as demonstrated in previous studies, highly agrees with the measured frequency response of brain tissue and can particularly capture the tissue's transition from a higher storage modulus at low frequencies to a higher loss modulus at high frequencies \citep{ruhland_combining_2026}. The fractional Kelvin-Voigt model is defined as
\begin{equation}\label{eq:fracKV}
   G^*(\omega)=\mu_e +\sum_{i=1}^2 c_i(i\omega)^{\alpha_i} 
\end{equation}
with the elastic shear modulus $\mu_e$, the springpot constant $c_i$ and powerlaw exponent $0\leq \alpha_i \leq 1$ of every springpot element. The springpot constant can be expressed in terms of the shear modulus $\mu_i$ and the viscosity $\eta_i$ 
\begin{equation}
    c_i = \mu_i^{1-\alpha_i}\eta_i^{\alpha_i}.
\end{equation}
Since $\mu_i$ and $\eta_i$ are linearly-dependent, we set $\eta = \SI{1}{\kilo\pascal \second}$ and thus obtain a single shear modulus for every springpot element \citep{klatt_viscoelastic_2010}. 
\begin{figure*}[h]
\centering
	\includegraphics{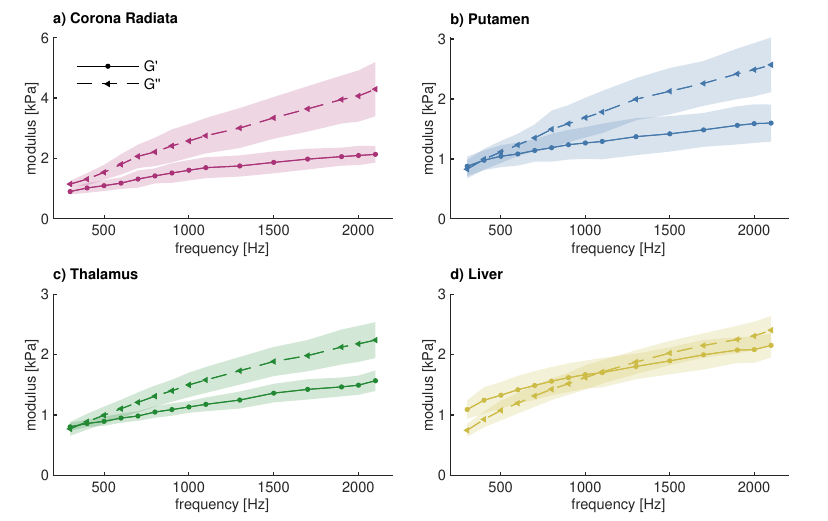}
	\caption{Measured storage (G') and loss moduli (G'') (mean $\pm$ standard deviation) for a) the corona radiata (n = 9), b) the putamen (n = 5), c) the thalamus (n = 5) and d) the liver (n = 5).}
	\label{fig:MREModu}
\end{figure*}
In the fractional Kelvin-Voigt model the spring defines elastic material behavior at \SI{0}{\hertz}. As the measured frequency response does not contain any information about the material response at low frequencies, we included a quasi static shear module as a storage modulus at \SI{0}{\hertz} to the frequency data for the brain tissue. The quasi static shear moduli were determined in multi modality rheometer experiments and are \SI{0.057}{\kilo\pascal} ($R = 1.1$) for the corona radiata, \SI{0.063}{\kilo\pascal} ($R = 0.68$) for the putamen and \SI{0.052}{\kilo\pascal} ($R = 0.58$) for the thalamus \citep{ruhland_combining_2026}. Since those values were determined, by calibrating the averaged measured response, we obtained a single value for the averaged material behavior. $R$ indicates the residuum of the calibration. For two fractional elements in parallel, the fractional element with the lower powerlaw exponent $\alpha$ dominates the lower frequency response and the fractional element with the higher exponent the high frequency response \citep{bonfanti_fractional_2020}. Therefore, we include an additional condition $\alpha_1 < \alpha_2$ in the calibration process and use the Nelder-Mead simplex algorithm with bounded variables and inequality constrains \texttt{fminsearcon} \citep{john_derrico_fminseachbnd_2025} in MATLAB Release 2023b for the model calibration. We identify the five free model parameters $\{\mu_e, \mu_1, \alpha_1, \mu_2, \alpha_2\}$ by minimizing the residuum
\begin{equation}\label{eq:res}
	R =\frac{\lVert\mathbf{x}^{meas}-\mathbf{x}^{calc}\rVert}{\lVert\mathbf{x}^{meas}\rVert}    
\end{equation} 
between the measured $\mathbf{x}^{meas}$ and the calculated data vector $\mathbf{x}^{calc}$.

\subsection{Statistical analysis}
All values are reported as median with the \SI{95}{\percent} bootstrap confidence interval. To identify the differences between the brain regions and the liver tissue, we performed statistical tests using the Statistics and Machine Learning Toolbox in MATLAB Release 2023b. The normal distribution of the parameters was examined by applying a Shapiro-Wilk test using the \texttt{swtest} function \citep{bensaida_shapiro-wilk_2025}. Since not all parameter sets are normally distributed, the non-parametric Kruskal-Wallis test was applied to determine significant differences between the groups. If a significant difference was observed in the group-wise comparison, the Wilcoxon rank sum test with the Holm-Bonferroni correction method to control the family-wise error rate was performed as a post-hoc test to identify the differences between the individual groups. For the Holm-Bonferroni correction, we used the \texttt{fwer\_holmbonf} function \citep{martinez-cagigal_multiple_2025}. To measure the correlation between the identified model parameters and the crossover frequency the Spearman's rank correlation coefficient was determined. For all tests a \textit{p}-value lower than 0.05 was considered to be significant.

\section{Results}\label{sec:results}
\subsection{Frequency response of the brain regions and the liver}\label{sec:MREmeasur}
Fig. \ref{fig:MREModu} shows the averaged storage and loss moduli together with the standard deviation for the corona radiata, the putamen, the thalamus and the liver measured with the tabletop MRE for frequencies from \SIrange{300}{2100}{\hertz}. Since the waves at \SI{300}{\hertz} occasionally suffered from an insufficient wave stimulation, resulting in nonphysical values for the storage and loss modulus, we executed some points of the \SI{300}{\hertz} data from the analysis. The resulting amount of data points at \SI{300}{\hertz} were $n_{CR} = 6$, $n_{P} = 5$,  $n_{T} =5$ and $n_{L} = 4$. The storage and loss moduli of the three brain regions and the liver demonstrate a continuous increase over the frequency range, with the loss modulus increasing at a faster rate than the storage modulus. Moreover, all tissues, except of the corona radiata, change from a dominating storage modulus at \SI{300}{\hertz} to a dominating loss modulus at \SI{2100}{\hertz}. However, the frequency at which the loss modulus exceeds the storage modulus differs for every tissue. On average, this frequency occurred the first time in the measurements at
\begin{figure*}[h]
\centering
	\includegraphics{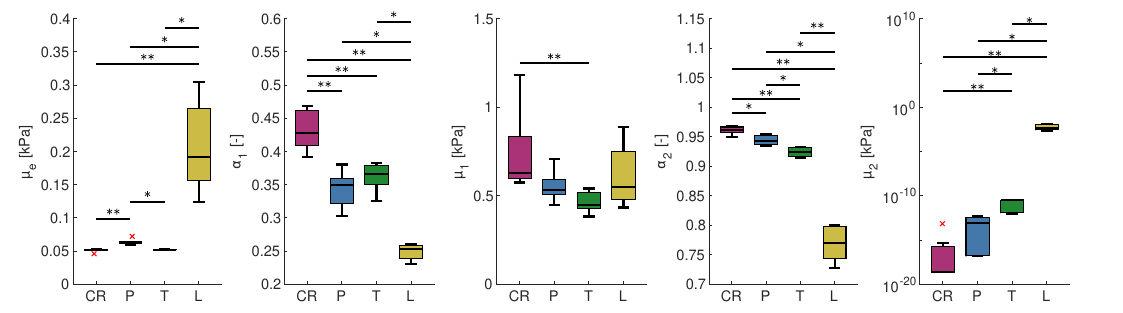}
	\caption{Comparison of the viscoelastic parameters for the corona radiata (CR), the putamen (P), the thalamus (T) and the liver (L). A fractional Kelvin-Voigt model with two fractional elements (shear modulus $\mu_i$ and powerlaw exponent $\alpha_i$) in parallel to an elastic spring (elastic shear modulus $\mu_e$) was calibrated based on the measured dynamic moduli. The box donated the median and the first and third interquartile range, the whiskers label the minimum and maximum data points and the outliers are given in red. Statistically significant differences between the groups are marked with ’*’ for p < 0.05 and ’**’ for p < 0.01.}
	\label{fig:MREPara}
\end{figure*}
\begin{figure}[h]
\centering
	\includegraphics{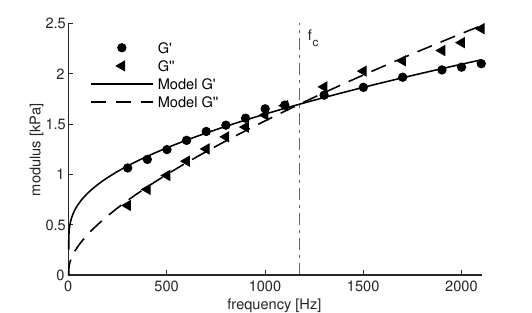}
	\caption{Crossover frequency $f_c$ of a representative liver sample. The crossover frequency is the frequency point, at which the storage and loss moduli intersect (G'($f_c$) = G''($f_c$)).}
	\label{fig:CrossData}
\end{figure}
\begin{figure}[h]
\centering
	\includegraphics{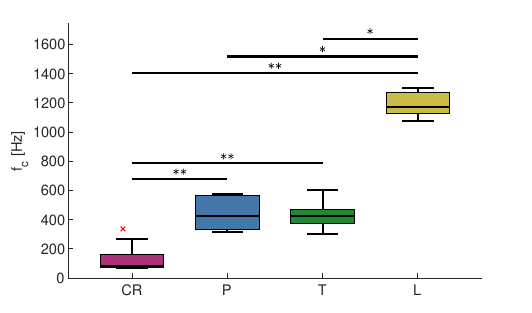}
	\caption{Comparison of the crossover frequency $f_c$ at which the loss modulus exceeds the storage modulus for the corona radiata (CR), the putamen (P),the thalamus (T) and the liver (L). The box donated the median and the first and third interquartile range, the whiskers label the minimum and maximum data points and the outliers are given in red. Statistically significant differences between the groups are marked with ’*’ for p < 0.05 and ’**’ for p < 0.01.}
	\label{fig:MRECross}
\end{figure}
 \SI{500}{\hertz} (\SI{95}{\percent}  CI: \qtyrange[range-phrase = {-}, range-units = single, number-unit-product = {}]{300}{600}{\hertz}) for the putamen, at \SI{500}{\hertz} (\SI{95}{\percent}  CI: \qtyrange[range-phrase = {-}, range-units = single, number-unit-product = {}]{400}{500}{\hertz}) for the thalamus and at \SI{1300}{\hertz} (\SI{95}{\percent}  CI: \qtyrange[range-phrase = {-}, range-units = single, number-unit-product = {}]{1100}{1300}{\hertz}) for the liver, indicating that this frequency is significantly higher for the liver than for the other two brain regions (see Tab. \ref{tab:pMREMeascross} for the \textit{p}-values). The corona radiata region displayed a greater loss than storage modulus over the whole frequency range.  

\subsection{Viscoelastic parameters for the brain regions and the liver}
Fig. \ref{fig:MREPara} compares the mechanical parameters of the fractional Kelvin-Voigt model for the corona radiata, the putamen, the thalamus and the liver and Tab. \ref{tab:pMREpara} provides the corresponding \textit{p}-values of the Kruskal-Wallis test. The comparison of the three brain regions revels that the putamen demonstrates the stiffest behavior at \SI{0}{\hertz}, represented by the elastic shear modulus $\mu_e$. For the first fractional element, we found a significantly higher first powerlaw exponent $\alpha_1$ in the corona radiata than in the other two regions. Moreover, the corona radiata exhibits a larger shear modulus $\mu_1$ than the thalamus. In the second fractional element, we observed significant differences between all three brain regions in both the powerlaw exponent $\alpha_2$ and the shear modulus $\mu_2$. The corona radiata exhibits the highest powerlaw exponent, followed by the putamen and then the thalamus. Regarding the shear modulus, the thalamus demonstrates the larges modulus of the three brain regions. Furthermore, a trend toward a higher modulus in the putamen relative to the corona radiata was observed (\textit{p} = 0.0599). Comparing the mechanical parameters of the brain regions to the liver, we observe substantial differences in all parameters expect of the shear modulus of the first fractional element. The elastic shear modulus and shear modulus of the second fractional element of the liver are considerably higher then those of all there brain regions. For the  powerlaw exponents of both fractional element, the liver tissue exhibits a smaller value than the brain tissue. Tab. \ref{tab:2fKV} contains the identified material parameters of the three brain regions and the liver.

\subsection{Crossover frequency of the brain regions and the liver}
Based on the calibrated material model, we determined the exact crossover point of the storage and loss modulus. Fig. \ref{fig:CrossData} shows the fractional Kelvin-Voigt model calibrated to the frequency response of one measurement with the identified crossover point. The crossover frequency was identified by finding 
\begin{figure*}[h]
\centering
	\includegraphics{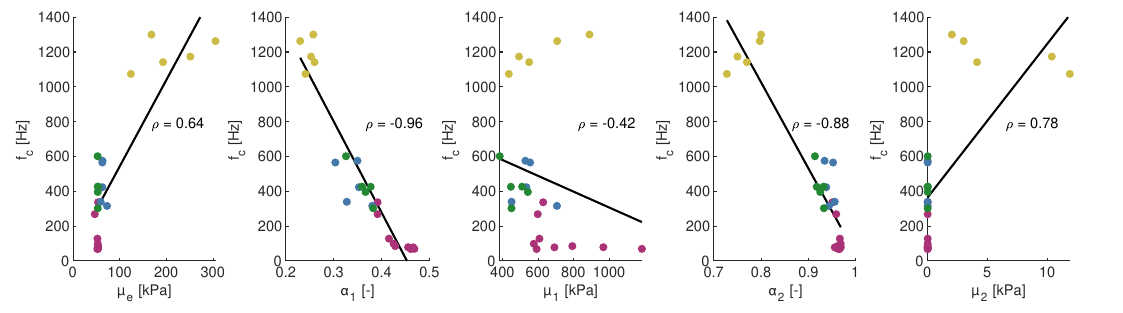}
	\caption{Correlation between the viscoelastic parameters of the fractional Kelvin-Voigt model and the crossover frequency $f_c$ for the corona radiata (magenta), the putamen (blue), the thalamus (green) and the liver (yellow). The corresponding correlation coefficients are given in Tab. \ref{tab:CorrCof}.}
	\label{fig:CrossCorr}
\end{figure*}
\begin{table*}[h]
	\centering
	\caption{Median (95 \% confidence interval) values of the parameters for the fractional Kelvin-Voigt model for the corona radiata (CR), the putamen (P), the thalamus (T) and the liver (L).}
	\renewcommand\arraystretch{1.5}
	\begin{tabular}{lccccc}
        \toprule
		& $\mu_e$ [kPa] & $\alpha_1$ [-] & $\mu_1$ [kPa]  & $\alpha_2$ [-] & $\mu_2$ [kPa]\\
		\midrule
		CR & 0.052 (0.052-0.053) &  0.429 (0.392-0.466) & 0.628 (0.597-0.967) & 0.962 (0.957-0.968) & 2.40e-19 (2.22e-19-4.83e-16)\\
 
		P & 0.062 (0.059-0.072)&  0.350 (0.303-0.381) & 0.534 (0.450-0.706) & 0.943 (0.934-0.955)& 7.96e-14 (1.61e-17-4.87e-13)\\

		T & 0.052 (0.052-0.053)&  0.366 (0.325-0.382)& 0.449 (0.383-0.542)& 0.924 (0.914-0.933)& 3.27e-11 (1.002-12-3.58e-11)\\

		L & 0.192 (0.123-0.305)&  0.252 (0.230-0.260)& 0.549 (0.434-0.889) & 0.770 (0.728-0.800)& 4.11e-3 (2.02e-3-1.18e-2)\\
	
	\end{tabular}
	\label{tab:2fKV}
\end{table*}
the intersection between the storage and the loss moduli. Fig. \ref{fig:MRECross} contains a comparison of the crossover frequencies determined from the calibrated fractional Kelvin-Voigt model model. The \textit{p}-values are provided in Tab. \ref{tab:pMREcross}. We observe significant differences in the crossover frequency between the corona radiata and the other two brain regions, as well as between all three brain regions and the liver. The corona radiata demonstrates the lowest crossover point at \SI{85}{\hertz} (\SI{95}{\percent}  CI: \qtyrange[range-phrase = {-}, range-units = single, number-unit-product = {}]{69}{269}{\hertz}), followed by the putamen \SI{423}{\hertz} (\SI{95}{\percent}  CI: \qtyrange[range-phrase = {-}, range-units = single, number-unit-product = {}]{316}{575}{\hertz}) and the thalamus \SI{426}{\hertz} (\SI{95}{\percent}  CI: \qtyrange[range-phrase = {-}, range-units = single, number-unit-product = {}]{302}{601}{\hertz}). The liver tissue exhibits the highest crossover frequency at \SI{1174}{\hertz} (\SI{95}{\percent}  CI: \qtyrange[range-phrase = {-}, range-units = single, number-unit-product = {}]{1074}{1300}{\hertz}) in the liver. For the putamen, the thalamus and the liver those values agree with the crossover of the measured storage and loss modulus in sec. \ref{sec:MREmeasur}. However, the model gives a more precise value for the crossover point, since we are not limited by the frequency interval used for the measurement. For the corona radiata, the crossover frequency is a predicted value, as this point lays outside the measured frequency range in the experiment. \\

To investigate the relation between the viscoelastic parameters and the crossover frequency, we performed a correlation analysis. Fig. \ref{fig:CrossCorr} demonstrates the correlation between the parameters of the fractional Kelvin-Voigt model and the crossover frequency and Tab. \ref{tab:CorrCof} contains the corresponding correlation coefficients and the \textit{p}-values. We found a strong negative correlation between both powerlaw exponents and the crossover frequency and a strong positive correlation between the shear modulus of the second fractional element and the crossover frequency. For the elastic shear modulus and the shear modulus of the first fractional element, the correlation to the crossover frequency was less pronounced.

\section{Discussion}\label{sec:discu}
We investigated the mechanical behavior of three porcine brain regions, i.e. the corona radiata, the putamen and the thalamus, and liver tissue at frequencies from \SIrange{300}{2100}{\hertz} with \textit{ex vivo} MRE. Our measured dynamic moduli for the brain regions are consistent with the values reported in previous \textit{ex vivo} studies on porcine corona radiata for frequencies from \SIrange{80}{140}{\hertz} 
\begin{table}[h]
	\centering
	\caption{Correlation coefficient and \textit{p}-value of the correlation analysis between the viscoelastic parameters of the fractional Kelvin-Voigt model and the crossover frequency.}
	\renewcommand\arraystretch{1.5}
	\begin{tabular}{lccccc}
        \toprule
		& $\mu_{e}$  & $\alpha_{1}$ & $\mu_{1}$ & $\alpha_{2}$ & $\mu_{2}$\\
		\midrule
		  $\rho$ & 0.64 & -0.96 & -0.42 & -0.88 & 0.78\\
          \textit{p} & 9.7e-4 & 1.6e-6 & 0.04 & 2.6e-6 & 1.0e-5
	\end{tabular}
	\label{tab:CorrCof}
\end{table}
\citep{vappou_magnetic_2007} and on porcine brain stem between \SI{300}{} and \SI{1500}{\hertz} \citep{braun_compact_2018}. Studies conducted on porcine brain regions different from the regions investigated in this study or on another species, reported dynamic moduli remarkably larger than ours \citep{riek_wide-range_2011, guertler_mechanical_2018}. Additionally, higher moduli were observed in \textit{in situ} measurements on porcine brain \citep{weickenmeier_brain_2018}. For the liver tissue, we obtained storage and loss moduli that align with moduli reported in the literature for calf liver in a frequency range from \SIrange{500}{1500}{\hertz} \citep{braun_compact_2018} and for porcine liver between \SI{500}{\hertz} and \SI{1000}{\hertz} \citep{ipek-ugay_tabletop_2015}. The storage moduli at \SI{0}{\hertz} for the three brain regions were taken from our previous study on porcine brain in the quasi-static domain. However, we did not experimentally characterize the quasi-static material behavior of the liver, and thus did not include a quasi-static storage modulus as an additional material parameter to the model calibration process for the liver. Still, the storage modulus we obtained for the liver at \SI{0}{\hertz}, which equals the elastic shear modulus $\mu_e$ of \SI{0.192}{\kilo\pascal}, is consistent with the values for porcine liver in the quasi-static or low frequency domain reported in the literature \citep{yang_mechanical_2020, mishra_rheological_2024}. Consequently, we can assume, that the calibrated fractional Kelvin-Voigt model for liver is valid over the whole frequency range.\\
In all tissues we observe a change from a dominating storage modulus to a dominating loss modulus with increasing frequencies. For a viscoelastic material, the storage modulus indicates the elastic material behavior, while the loss modulus reflects the viscose behavior. In previous studies of brain tissue a more elastic material behavior, thus a higher storage than loss modulus, was reported in quasi-static rheometer tests \citep{ruhland_combining_2026}, in low frequency oscillating rheometry \citep{vappou_magnetic_2007, mishra_rheological_2024} and with MRE for frequencies up to \SI{600}{\hertz} \citep{vappou_magnetic_2007, riek_wide-range_2011, guertler_mechanical_2018}. Similarly, a more elastic material behavior was reported for liver at low frequencies with oscillating rheometry \citep{klatt_viscoelastic_2010, mishra_rheological_2024} and for frequencies up to \SI{1000}{\hertz} with MRE \citep{riek_wide-range_2011, ipek-ugay_tabletop_2015}. In this study, we obtained a more elastic material response for brain tissue for frequencies up to approximately \SI{500}{\hertz}, depending on the brain region, and for the liver tissue until approximately \SI{1200}{\hertz}. At high frequencies both the brain and the liver tissue behaved more viscous (Fig. \ref{fig:MREModu}). This indicates that brain and liver tissue transits from an elasticity-dominated material behavior at low frequencies to a viscosity-dominated behavior at high frequencies. The frequency point at which this behavior changes, and thus the loss modulus crosses the storage modulus, is the crossover frequency $f_c$. We determined the crossover frequencies for every brain region and the liver using the calibrated fractional Kelvin-Voigt model (Fig. \ref{fig:MRECross}). Interestingly, we observe similar differences in the crossover frequencies as in the viscoelastic parameters between the tissues. A correlation analysis revealed a relation between the viscoelastic parameters and the crossover frequencies (Fig. \ref{fig:CrossCorr}). Therefore, we conclude that the crossover frequency is an indicator of the viscoelastic behavior of the material, distinguishing between different brain regions, as well as between the brain and liver tissue. To identify the crossover frequencies, we used the storage and loss modulus curves derived from the calibrated fractional Kelvin-Voigt model, allowing for an exact determination of the crossover frequency. The identification based on the experimentally determined storage and loss moduli is restricted by the interval between the measurement points. In this study, the interval was \SI{100}{\hertz} for the lower frequencies and \SI{200}{\hertz} for the higher frequencies. Therefore, the crossover frequencies obtained with the measured moduli were less precise than those obtained with the modeled curves. However, the experimentally determined crossover frequencies (Sec. \ref{sec:MREmeasur}) align with the exact crossover frequencies (Fig. \ref{fig:MRECross}). Consequently, the crossover frequency can be determined without calibrating a viscoelastic material model to the measured storage and loss modulus curves. More precise values for the crossover frequency can be achieved by reducing the distance between the measurement points.

\section{Conclusion}\label{sec:concl}
In this study, we investigated the viscoelastic behavior of three porcine brain regions, i.e. the corona radiata, the putamen and the thalamus, and porcine liver in the high-frequency domain. For all tissues, we observed a change in the viscoelastic behavior from a higher elasticity at lower frequencies to a dominating viscosity at high frequencies. This transition in material behavior is marked by the crossover frequency $f_c$, the frequency at which the loss modulus exceeds the storage modulus. We demonstrated that the crossover frequency correlated to the viscoelastic behavior and reflects the regional differences in the brain, as well as the mechanical differences between brain and liver tissue. Moreover, we showed that the crossover frequency can directly be derived from the measured dynamic moduli, and thus is independent of an appropriate material model and fitting strategy selection to characterize the tissue's viscoelastic behavior. The crossover frequency is therefore a practical, model-independent constant for the rapid viscoelastic characterization of biological tissue in elastography.

\section*{Acknowledgments}
This work was funded by the German Research Foundation (DFG) project 460333672 CRC1540 EBM, CRC1340 Matrix-in-vision, FOR5628, 540759292 Sa901/33-1 M5.

\bibliographystyle{elsarticle-harv} 
\bibliography{references}
  
\appendix
\section{Result of the statistical analysis}\label{app1}

See Tab. \ref{tab:pMREpara} to \ref{tab:pMREcross}.

\begin{table}[H]
	\centering
	\caption{Kruskal-Wallis and Wilcoxon \textit{p}-value for the parameters of the fractional Kelvin-Voigt model}
	\begin{tabular}{lccccc}
        \toprule
            \multicolumn{3}{l}{Kruskal-Wallis} & \multicolumn{3}{l}{\textit{p} = 4.22e-4}\\
        \midrule
         $\mu_0$&  & CR & P & T & L \\
		        &  CR & - & 5.99e-3 & 0.79 & 5.00e-3 \\
                & P &  & - & 0.03 & 0.02 \\
                & T &  &  & - & 0.02 \\
                & L &  &  &  & - \\
        \midrule
            \multicolumn{3}{l}{Kruskal-Wallis} & \multicolumn{3}{l}{\textit{p} = 1.55e-4}\\
        \midrule
         $\alpha_1$&  & CR & P & T & L \\
		        &  CR & - & 5.99e-3 & 5.00e-3 & 4.00e-3 \\
                & P &  & - & 0.31 & 0.02 \\
                & T &  &  & - & 0.02 \\
                & L &  &  &  & - \\
        \midrule
            \multicolumn{3}{l}{Kruskal-Wallis} & \multicolumn{3}{l}{\textit{p} = 0.01}\\
        \midrule
         $\mu_1$&  & CR & P & T & L \\
		        &  CR & - & 0.14 & 5.99e-3 & 0.48 \\
                & P &  & - & 0.38 & 0.84 \\
                & T &  &  & - & 0.67 \\
                & L &  &  &  & - \\
        \midrule
            \multicolumn{3}{l}{Kruskal-Wallis} & \multicolumn{3}{l}{\textit{p} = 1.39e-4}\\
        \midrule
         $\alpha_2$&  & CR & P & T & L \\
		        &  CR & - & 0.03 & 5.99e-3 & 5.00e-3 \\
                & P &  & - & 0.02 & 0.02 \\
                & T &  &  & - & 7.93e-3 \\
                & L &  &  &  & - \\
        \midrule
            \multicolumn{3}{l}{Kruskal-Wallis} & \multicolumn{3}{l}{\textit{p} = 2.31e-4}\\
        \midrule
         $\mu_2$&  & CR & P & T & L \\
		        &  CR & - & 0.06 & 5.99e-3 & 5.00e-3 \\
                & P &  & - & 0.03 & 0.02 \\
                & T &  &  & - &  0.02 \\
                & L &  &  &  & - \\
	\end{tabular}
	\label{tab:pMREpara}
\end{table}

\begin{table}[H]
	\centering
	\caption{Kruskal-Wallis and Wilcoxon \textit{p}-value for the measured crossover frequency}
	\begin{tabular}{lcccc}
        \toprule
        \multicolumn{3}{l}{Kruskal-Wallis} & \multicolumn{2}{l}{\textit{p} = 7.18e-3}\\
        \midrule
         $f_c$& & P & T & L \\
		       &  P & - & 1 & 0.02  \\
                & T &  & - & 0.02 \\
                & L &  &  & - \\
        \midrule
	\end{tabular}
	\label{tab:pMREMeascross}
\end{table}

\begin{table}[H]
	\centering
	\caption{Kruskal-Wallis and Wilcoxon \textit{p}-value for the crossover frequency of the fractional Kelvin-Voigt model}
	\begin{tabular}{lccccc}
        \toprule
        \multicolumn{3}{l}{Kruskal-Wallis} & \multicolumn{3}{l}{\textit{p} = 2.48e-4}\\
        \midrule
         $f_c$&  & CR & P & T & L \\
		        &  CR & - & 9.99e-3 & 7.99e-3 & 5.99e-3 \\
                & P &  & - & 1 & 0.02 \\
                & T &  &  & - & 0.02 \\
                & L &  &  &  & - \\
        \midrule
	\end{tabular}
	\label{tab:pMREcross}
\end{table}

\end{document}